\documentstyle[12pt]{article}
\begin{document}
\begin{titlepage}

\hfill{July 1997}

\hfill{UM-P-97/48}

\hfill{RCHEP-97/08}

\vskip 0.9 cm

\centerline{{\large \bf 
An alternative $SU(4) \otimes SU(2)_L \otimes SU(2)_R$ model}}
\vskip 1.3 cm
\centerline{R. Foot}

\vskip 1.0 cm
\noindent
\centerline{{\it School of Physics}}
\centerline{{\it Research Centre for High Energy Physics}}
\centerline{{\it The University of Melbourne}}
\centerline{{\it Parkville 3052 Australia }}

\vskip 2.0cm

\centerline{Abstract}
A simple alternative to the usual Pati-Salam model is
proposed. The model allows quarks and leptons to be unified
with gauge group $SU(4) \otimes SU(2)_L \otimes SU(2)_R$
at a remarkably low scale of about 1 TeV.
Neutrino masses in the model arise radiatively and
are naturally light.
\vskip 0.7cm
\noindent

\end{titlepage}

\vskip 1.2cm

In the standard $SU(2)_L \otimes U(1)_Y$ model
of electroweak interactions\cite{bp} quarks and 
leptons have many similarities. For example, 
there are three generations of quarks and three 
generations of leptons,
with 2 quarks and 2 leptons in each generation.
Furthermore both left-handed quarks and left-handed leptons
transform as $SU(2)_L$ doublets while both right-handed quarks
and right-handed leptons are $SU(2)_L$ singlets.

The similarity of quarks and leptons may be due to a
spontaneously broken exact symmetry. This symmetry
if it exists could be either continuous\cite{ps} 
or discrete\cite{fl}. In the case where it is discrete, 
unification of quarks and leptons can occur at 
very low scales of around a TeV\cite{fl,flv}.
If quarks and leptons are related by a 
continuous Pati-Salam type gauge symmetry
it is usually assumed that this symmetry is broken 
at a very high scale $M \ge 10^{11}\ GeV$ which means 
that the idea cannot be tested directly in any 
conceivable experiment.  However if there is no 
left-right symmetry then it is nevertheless still 
possible that the standard model is a remnant of a 
gauge model with Pati-Salam gauge symmetry
broken at a relatively low scale of 
$1000\ TeV$ or even less\cite{km, vw,volkas}.
The purpose of this letter is to point out that there 
exists an interesting alternative Pati-Salam type model 
which can be broken at a much lower scale of the order 
of a TeV.

The gauge symmetry of the model is 
\begin{equation}
SU(4) \otimes SU(2)_L \otimes SU(2)_R.
\label{1}
\end{equation}
Under this gauge symmetry the fermions of each generation transform 
in the anomaly free representations:
\begin{equation}
Q_L \sim (4,2,1),\  Q_R \sim (4, 1, 2), \ f_L \sim (1,2,2).
\label{2}
\end{equation}
The minimal choice of scalar multiplets which can both
break the gauge symmetry correctly and give all of the 
charged fermions mass is
\begin{equation}
\chi_L \sim (4, 2, 1), \ \chi_R \sim (4, 1, 2),\ \phi \sim (1,2,2).
\label{3}
\end{equation}
These scalars couple to the fermions as follows:
\begin{equation}
{\cal L} = \lambda_1 \bar Q_L (f_L)^c \tau_2 \chi_R 
+ \lambda_2 \bar Q_R f_L \tau_2 \chi_L 
+ \lambda_3 \bar Q_L \phi \tau_2 Q_R  + 
\lambda_4 \bar Q_L \phi^c \tau_2 Q_R  
+ H.c.,
\label{4}
\end{equation}
where the generation index has been suppressed and $\phi^c = \tau_2 
\phi^* \tau_2$.  Under the $SU(3)_c \otimes U(1)_T$ 
subgroup of $SU(4)$, the $4$ representation has the branching rule,
$4 = 3(1/3) + 1(-1)$.  We will assume that the 
$T=-1, I_{3R} = -1/2 \ (I_{3L}=1/2)$ components of $\chi_{R} (\chi_L)$
gain non-zero Vacuum Expectation Values (VEVs) as well as 
the $I_{3L} = I_{3R} = -1/2$ and $I_{3L} = I_{3R} = 1/2$
components of the $\phi$.
We denote these VEVs by $w_{R,L}, u_{1,2}$ respectively.
In other words,
\begin{eqnarray}
\langle \chi_R (T = -1, I_{3R} = -1/2) \rangle = w_R, \
\langle \chi_L (T = -1, I_{3L} = 1/2) \rangle = w_L, 
\nonumber \\
\langle \phi (I_{3L} = I_{3R} = -1/2)\rangle = u_1,\
\langle \phi (I_{3L} = I_{3R} = 1/2)\rangle = u_2.
\end{eqnarray}
We will assume that the VEVs satisfy
$w_R > u_{1,2}, w_L$
so that the symmetry is broken as follows:
\begin{eqnarray}
&SU(4)\otimes  SU(2)_L \otimes SU(2)_R&
 \nonumber \\
&\downarrow \langle \chi_R \rangle&
\nonumber \\
&SU(3)_c \otimes SU(2)_L \otimes U(1)_Y &
\nonumber \\ 
&\downarrow \langle \phi \rangle, \langle \chi_L \rangle
\nonumber \\
&SU(3)_c \otimes U(1)_Q&
\end{eqnarray}
where $Y = T -2I_{3R}$ is the linear combination of 
$T$ and $I_{3R}$ which annihilates $\langle \chi_R \rangle$ 
(i.e. $Y\langle \chi_R \rangle = 0$).  Observe that in 
the limit where $w_R \gg w_L, u_1, u_2$, the model 
reduces to the standard model. The VEV $w_R$ breaks the 
gauge symmetry to the standard model subgroup. This VEV 
also gives large (electroweak invariant) masses to an 
$SU(2)_L$ doublet of exotic fermions, which  
have electric charges $-1, 0$.  We will denote these 
exotic fermions with the notation $E^-, E^0$.
These exotic fermions must have masses greater than 
$M_Z/2$ otherwise they would contribute to the $Z$ width.
Observe that the right-handed chiral components of 
the usual charged leptons are contained in $Q_R$. 
They are the $T=-1, I_{3R} = 1/2$ components.
The usual left-handed leptons are contained in 
$f_L$ along with the right-handed components 
(CP conjugated) of $E^0, E^-$.
It is instructive to write out the fermion multiplets
explicitly. For the first generation,
\begin{eqnarray}
Q_L = \left(\begin{array}{cc}
d_y & u_y   \\
d_g & u_g  \\
d_b & u_b  \\
E^- & E^0 
\end{array}
\right)_L,\
Q_R = \left(\begin{array}{cc}
u_y & d_y  \\
u_g & d_g \\
u_b & d_b \\
\nu & e  
\end{array}
\right)_R, \
f_L = \left(\begin{array}{cc}
(E^-_R)^c & \nu_L \\
(E^0_R)^c & e_L 
\end{array}
\right),
\end{eqnarray}
and similarly for the second and third generations.
In the above matrices the first column of $Q_L$ $(f_L,\ Q_R)$
is the $I_{3L} (I_{3R}) = -1/2$ component while the second
column is the $I_{3L} (I_{3R}) = 1/2$ component.
The four rows of $Q_L, Q_R$ are the four colours and
the rows of $f_L$ are the $I_{3L} = \pm 1/2$ components.
Observe that the VEVs $w_L, u_{1,2}$ have the quantum 
numbers $I_{3L} = -1/2, Y = 1$
(or equivalently $I_{3L} = 1/2, Y = -1$).
This means that the standard model subgroup, 
$SU(3)_c \otimes SU(2)_L \otimes U(1)_Y$
is broken to $SU(3)_c \otimes U(1)_Q$ in the usual way (with
$Q = I_{3L} + Y/2$). 

Having established that the model is a phenomenologically viable
extension to the standard model, we now comment on 
various features of the model.

Observe that the model has the rather unusual feature that
the scalar multiplets required to break the gauge symmetry
and give the fermions masses have precisely the same quantum
numbers as the fermion multiplets of a generation [compare
Eq.(\ref{2}) and Eq.(\ref{3})].

In the model the ordinary neutrinos are naturally light. 
The neutrino masses vanish at tree-level given
the particle content of the theory. The model does however
have a light singlet neutrino, $\nu_R$.
This electroweak singlet occupies the
$T =-1, I_{3R} = -1/2$ component of the $Q_R$ multiplet.
Note however that with the minimal Higgs content, this field 
cannot couple to the ordinary left-handed neutrinos 
(at tree level).
Observe that Majorana neutrino masses arise from the $W_{L,R}$
gauge interactions at the one
loop level. Assuming diagonal couplings and examining  
$\nu_e$ for definiteness, the Feynman
diagram for the $\nu_e$ mass is shown in Figure 1.
Calculating this finite 1-loop diagram we find that
\begin{equation}
m_{\nu} \simeq {2g_R g_L \over (4\pi)^2} \left[
{g_R g_L u_1 u_2 \over M^2_{W_R}}\right]
\left[ {m_e m_d M_E \over M_E^2 - M_{W_L}^2}\right]
log\left({M_E^2 \over M_{W_L}^2}\right),
\end{equation}
where $g_{L,R}$ are the $SU(2)_{L,R}$ gauge coupling constants
and we have assumed that $M_E^2 \ll M_{W_R}^2$. Clearly the
neutrino masses are naturally light given that $M_{W_R},
M_E \gg m_e, m_d$.

The gauge interactions of the model conserve an 
unbroken baryon number symmetry. This baryon charge is
defined as $B = B' + T$ where the $B'$ charges of 
$Q_L, Q_R, \chi_{L,R}$ are $1$ and the $B'$ charges of 
$f_L, \phi$ are $0$. The existence of the baryon number 
symmetry implies that protons and neutrons are absolutely 
stable in the model.

Because the right-handed charged leptons belong to the
same multiplet as the right-handed quarks
there will be gauge interactions of the form
\begin{equation}
{\cal L} = {g_s \over \sqrt{2}}\bar D_R^i 
W'_{\mu}\gamma^{\mu} K'^{ij} l_R^j + H.c.,
\end{equation}
where $i,j = 1,...,3$ are family indices, that is $D_R^1 = d_R,
D_R^2 = s_R, D_R^3 = b_R$, $l_R^1 = e_R, l_R^2 = \mu_R,
l_R^3 = \tau_R$ and $W'_{\mu}$ are electrically charged 
$2/3$ gauge bosons (which gain masses from $\chi_R$ at the 
first step of symmetry breaking).
The matrix $K'$ is a Cabbibo-Kobayashi-Maskawa type matrix.
The most stringent bound on the symmetry breaking scale $w_R$
is expected to arise from $K_L \to \mu^{\pm} e^{\mp}$ decays.
The decay $\bar K^0 \to \mu^- e^+$ arises from a Feynman
diagram with a T-channel exchange of a $W'$ gauge boson.
This diagram corresponds (after a Fierz rearrangement)
to the effective four fermion Lagrangian density,
\begin{equation}
{\cal L}_{eff} = {G_X \over \sqrt{2}}
\bar d\gamma_{\mu} (1 + \gamma_5)s 
\bar \mu \gamma^{\mu}(1 + \gamma_5) e
+ H.c.,
\end{equation}
where $G_X = \sqrt{2}g_s^2(M_{W'})/8M^2_{W'}$.
Using this effective Lagrangian it is straightforward to
calculate the decay rate. We find,
\begin{equation}
\Gamma (\bar K^0 \to \mu^-  e^+) \simeq {G_X^2
f_K^2 \over 8\pi} M_K m_{\mu}^2,
\label{ii}
\end{equation}
where $f_K$ is the $K$ meson decay constant, $M_K$
is the $K$ meson mass and we have assumed that the
mixing matrix $K'_{ij}$ is approximately diagonal. 
Evaluating the above equation we find that
\begin{equation}
Br(\bar K^0 \to \mu^-  e^+) \simeq 
10^{-2} \left( {TeV \over M_{W'}}\right)^4.
\end{equation}
The current experimental bound,
$Br(\bar K^0 \to \mu^{\pm} e^{\mp}) < 
3.3 \times 10^{-11}$\cite{pdg}, implies the limit
\begin{equation}
M_{W'} \stackrel{>}{\sim} 140 \ TeV,
\label{100TeV}
\end{equation}
assuming that 
the mixing matrix $K'^{ij}$ is approximately 
diagonal.  This bound is the 
most stringent bound on the model in the case where 
$K'^{ij}$ is diagonal.  However in the model there is no 
relationship between the
charged lepton mass matrix and the quark masses. Indeed, in
the model they are proportional to the VEVs of different
scalar multiplets and have independent Yukawa couplings.  
One consequence of this is that the mixing matrix $K'^{ij}$ 
connecting the right-handed quarks with the right-handed leptons
is theoretically unconstrained (except of course, for 
the unitrary requirement).
For example, the $W'$ could couple
$s_R$ predominately with $\tau_R$ (this possibility was discussed
in the context of the usual Pati-Salam model in Ref.\cite{vw}).
If this is the case then $K_L$ decays do not give any stringent 
bounds on the model.  
In order to explore this scenario further, we will assume 
for definiteness that $W'$ couples $s_R$ with $\tau_R$,
$d_R$ with $e_R$ and $b_R$ with $\mu_R$. This corresponds
to a $K'^{ij}$ matrix of the form
\begin{eqnarray}
K'= \left(
\begin{array}{ccc}
1&0&0 \\
0&0&1 \\
0&1&0 
\end{array}
\right).
\label{zz}
\end{eqnarray}
Of course it would not seem natural for the zero elements 
of this mixing matrix to be exactly zero. However we will
assume that they are zero to illustrate a point.
In the case of the usual Pati-Salam model with the anzatz
Eq.(\ref{zz}), the most stringent bound on $M_{W'}$ 
comes from the $W'$ contribution to the 
decay $\pi^+ \to e^+ \nu_L$\cite{vw}.
This process leads to a quite stringent bound of
$M_{W'} \stackrel{>}{\sim} 250 \ TeV$\cite{vw} for that model.
This bound arises by calculating the interference term 
between the amplitudes arising from the standard 
model contribution and the Pati-Salam contribution.
However the decay $\pi^+ \to e^+ \nu$ does not provide 
a stringent constraint for the alternative Pati-Salam model.
There are two reasons for this. First, the $W'$
mediates the decay $\pi^+ \to e^+ \nu_R$ (rather
than $\pi^+ \to e^+ \nu_L$).
Because the final state is distinct from the standard model
process $\pi^+ \to e^+ \nu_L$,  
there will obviously be no interference term between the amplitudes
of the two processes. Second, the $W'$ of the alternative model
only couples to right-handed quarks and leptons.  
This means that the decay $\pi^+ \to e^+ \nu_R$ is 
helicity suppressed by a factor $m_e^2/m^2_{\pi}$ (which
is also the case for the standard model contribution).
In fact,
\begin{equation}
{\Gamma (\pi^+ \to e^+ \nu_R) \over
\Gamma(\pi^+ \to e^+ \nu_L)} \simeq {G_X^2 \over G_F^2},
\end{equation}
where $G_F$ is the
usual Fermi constant. The above contribution to $\pi^+ \to
e^+ \nu$ decay leads to a violation of lepton universality and 
implies a small modification to the ratio $R$ where
$R \equiv \Gamma (\pi^+ \to e^+ \nu)/\Gamma (\pi^+ \to
\mu^+ \nu)$.  Using $\alpha_s (M_{W'}) \sim 1/10$, 
we find,
\begin{equation}
{\delta R \over R} 
\simeq 4 \times 10^{-4}\left( {TeV \over M_{W'}}\right)^4.
\label{ew}
\end{equation}
The theoretical prediction for $R$ agrees within errors to the 
experimental measurement and thus Eq.(\ref{ew}) 
can be compared to the experimental error
$\delta R/R \sim 0.003$\cite{pdg}. Clearly then, $\pi^+$ decay does
not lead to any significant bound for the model.

A more stringent bound on the model
[assuming the ansatz Eq.(\ref{zz})]
arises from the $W'$ mediated
rare $B_d^0$ decay, $\bar B_d^0 \to \mu^- e^+$.
In the case of the usual Pati-Salam model with the ansatz
Eq.(\ref{zz}), the bound $M_{W'} \stackrel{>}{\sim} 16 \ TeV$ 
was derived in Ref.\cite{vw}.  However in my model the 
bound arising from this process is much less stringent.
The main difference is that the $W'$ of the usual 
Pati-Salam model couples vectorially where as the $W'$ of 
the alternative Pati-Salam model couples only
to right-handed quarks and leptons and $\bar B_d^0$ 
decays will be helicity suppressed by a factor 
$\sim m^2_{\mu}/M^2_B$.  In fact the width for the
decay $\bar B_d^0 \to \mu^- e^+$ [assuming the 
ansatz Eq.(\ref{zz})] is given by Eq.(\ref{ii})
with the replacement $K \to B$.
Evaluating the resulting equation we find that
\begin{equation}
Br(\bar B_d^0 \to \mu^-  e^+) \simeq 
3\times 10^{-6} \left( {TeV \over M_{W'}}\right)^4.
\end{equation}
The current experimental bound,
$Br(\bar B_d^0 \to \mu^{\pm} e^{\mp}) < 
6 \times 10^{-6}$\cite{pdg}, implies the limit
\begin{equation}
M_{W'} \stackrel{>}{\sim} 800 \ GeV.
\label{1TeV}
\end{equation}
If the $W'$
gauge boson is light, then how large can the zero elements
of $K'_{ij}$ be? The most stringently constrained element
is the $K'_{s\mu}$ entry [which is the $K'_{22}$ element of
Eq.(\ref{zz})]. This entry is constrained to be
$K'_{s\mu} \stackrel{<}{\sim} 10^{-4}$ if $M_{W'}\simeq 1\ TeV$ 
given the experimental bound $Br(K_L \to \mu^{\pm}e^{\mp})
< 3.3 \times 10^{-11}$\cite{pdg}.
 
Note that so-called vector lepto-quarks have been studied
which have similar properties to the $W'$ gauge bosons\cite{ab}.
However it is usually assumed that vector lepto-quarks must
couple to only one generation if they are to be light enough
to be seen in collider experiments. One result of this paper
is that it is possible to have light vector lepto-quarks
coupling chirally to all three quarks and leptons.
Furthermore, the model provides a concrete renormalizable
framework where vector lepto-quarks with chiral couplings
are gauge fields and may thus be fundamental particles.

In addition to the exotic $W'$ gauge bosons, the model 
contains $W_R^{\pm}$ and $Z'$ gauge bosons.
The exotic gauge boson mass matrix 
arises from the Lagrangian density terms
(in the limit where $w_R \gg w_L, u_{1,2}$) 
\begin{equation}
{\cal L} = \left(D_{\mu} \langle \chi_R \rangle\right)^{\dagger} 
D^{\mu} \langle \chi_R \rangle.
\label{ee}
\end{equation}
where the covariant derivative is given by
\begin{equation}
D_{\mu} = \partial_{\mu} + ig_s G_{\mu}^a \Lambda_a + 
ig_L W_{L\mu}^i \tau^i_L/2 + ig_R W_{R\mu}^i \tau^i_R/2,
\end{equation} 
where $a = 1,...,15, i=1,...,3$ and $G_{\mu}^a$, $W_{L\mu}$, 
$W_{R\mu}$ ($\Lambda_a$, $\tau^i_L/2$, $\tau^i_R/2$) are the 
$SU(4)$, $SU(2)_L$, $SU(2)_R$ gauge bosons (generators) 
respectively. By examining the exotic gauge boson mass 
matrix Eq.(\ref{ee}), it is possible to obtain the 
usual weak mixing angle, $\sin\theta_w \equiv e/g_L$, as
a function of the couplings $g_s, g_L, g_R$.
We find that
\begin{equation}
\sin^2 \theta_w (M_{W'}) = 
{g_s^2 (M_{W'}) g_R^2 (M_{W'}) \over 
g_s^2 (M_{W'}) g_R^2 (M_{W'}) +  
g_s^2 (M_{W'}) g_L^2 (M_{W'}) +  
{2 \over 3}g_L^2 (M_{W'}) g_R^2 (M_{W'})}.  
\label{jj}
\end{equation}
Assuming that $\sin^2 \theta_w (M_{W'}) \simeq 1/4$ which
is appropriate for $M_{W'} \sim  1 \ TeV$, 
Eq.(\ref{jj}) implies that $g_R (M_{W'}) \simeq 
g_L(M_{W'})/\sqrt{3}$.  Also it is easy to show that
$M_{W'} \simeq \sqrt{2/3}M_{Z'} \simeq (g_s/g_R)M_{W_R}$.
Furthermore the $Z'$ gauge boson couples to fermions via the
interaction Lagrangian density,
\begin{equation}
{\cal L} = -g_s Z'_{\mu} J^{\mu},
\end{equation}
where the current is given by
\begin{equation}
J^{\mu} = \bar \psi Q' \gamma^{\mu} \psi.
\end{equation}
In the above equation, the summation of fermion fields is implied
and the generator $Q'$ is given approximately by
\begin{equation}
Q'\simeq \sqrt{{3 \over 8}}\left( T + 
{4\over 9}{g_L^2 \over g_s^2}I_{3R} \right),
\end{equation}
where we have again assumed that $\sin^2 \theta_w (M_{W'}) \simeq 1/4$.
From the contributions of the $Z', W_R^{\pm}$ to low energy experiments,
a limit of $M_{Z'}, M_{W_R} \stackrel{>}{\sim} 
0.5 - 1 \ TeV$ is expected\cite{rizzo}.
Thus, we argue that the model is quite weakly constrained given
that the exotic symmetry breaking scale can be as low as a TeV or so.

The model contains scalar lepto-quarks which could
be relatively light (e.g. a few hundred GeV).
In particular $\chi_L$ contains $SU(3)_c$ triplet $SU(2)_L$
doublet scalars coupling the left-handed leptons with the right
handed $d$ type quarks.  From Eq.(\ref{4}), we can deduce that 
\begin{equation}
{\cal L}_{\chi} =  \lambda_2 \bar d_R L_L \chi + H.c.,
\label{y}
\end{equation}
where $\chi$ is the colour triplet component of $\chi_L$,
and $L_L = (\nu_L, \ e_L)^T$.  Note that $\chi$-type
lepto-quarks coupling to $d_R$ with masses of around 
$200$ GeV have been put forward as a possible explanation of the 
excess high $Q^2$ Hera events\cite{hr}.
(However, the Hera anomaly is only a 2-3 sigma excess
and may disappear when more data is taken).

Observe that the Yukawa Lagrangian density, Eq.(\ref{4})
implies that the right-handed charged leptons
will mix slightly with the $E^-$ exotic fermions.
For one generation, the mixing has the form,
\begin{eqnarray}
{\cal L}_{mass} = 
(\bar e_L \bar E_L)\left(
\begin{array}{cc}
\lambda_2 w_L & 0 \\
\lambda_3 u_1 + \lambda_4 u_2 & \lambda_1 w_R 
\end{array}
\right)
\left( \begin{array}{c}
e_R \\
E_R
\end{array} \right) + H.c.
\end{eqnarray}
Note however that this mixing is expected to be small because 
$u_{1,2} \ll w_R$ [given the bound Eq.(\ref{1TeV})].
Because the exotic $E^-$ fermions do not have canonical 
$Z$ couplings, the mixing will induce small flavour changing 
neutral current (FCNC) couplings in the general case 
of three generation mixing. The mixing will be constrained 
by processes such as $\mu^- \to e^+ e^- e^-$.

Finally, I would like to comment briefly on a cosmological
issue.  Within the context of the standard big bang model,
the phase transition at the temperature scale
$T \sim w_R$ will generate monopoles.
Monopoles occur because a semi-simple gauge theory
is broken down to a gauge symmetry with a $U(1)$ factor.
Monopoles in the early Universe can
be a problem if they are too abundant.
This issue has been examined in Ref.\cite{volkas}
for the case of the usual Pati-Salam model broken
at a low scale. It was concluded that there is no problem
if the symmetry breaking scale is low, which is the case 
that is being considered in the present paper.

In conclusion, an alternative Pati-Salam type gauge model has
been proposed. The model allows quarks and leptons to be unified
with gauge group $SU(4) \otimes SU(2)_L \otimes SU(2)_R$
at a relatively low scale. We argue that present data
does not constrain this model very stringently (c.f.
the usual Pati-Salam model). As a
consequence the exotic gauge boson masses (and thus the 
symmetry breaking scale) can be as low as about $1$ TeV. 
Neutrino masses arise radiatively in the model and
are naturally light.
\newpage
\vskip 0.8cm
\noindent
{\bf Acknowledgements}
\vskip 0.4cm
\noindent
The author would like to thank Ray Volkas for the usual discussions 
and for pointing our some relevant papers.
He would also like to thank X-G. He and J.P. Ma for a discussion
and to J.Bowes for help with the Figure.
The author is supported by an Australian Research Fellowship.

\newpage
\vskip 1cm
\noindent
{\bf Figure Caption}
\vskip 0.5cm
\noindent
Figure 1: 
1-loop Feynman diagram which leads to small  
electron neutrino Majorana mass. 
There will be similar diagrams for the other neutrinos.
(The $W_L W_R$ mixing mass squared is obtained from
${\cal L} = (D_{\mu} \langle \phi \rangle )^{\dagger} 
D^{\mu} \langle \phi \rangle$ and is given 
by $\mu^2 = g_R g_L u_1 u_2$).

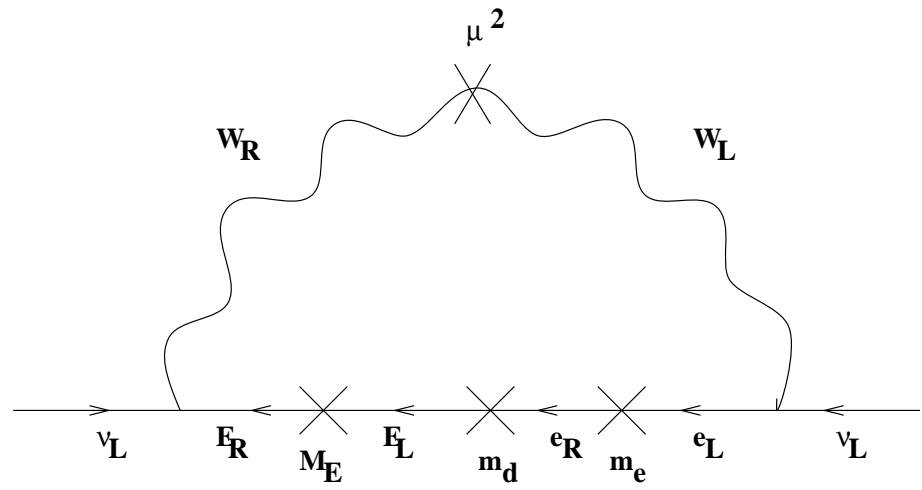
\end{document}